\documentstyle[amssymb,preprint,aps]{revtex}
\tightenlines
\begin{document}
\draft
\title{ Quantum coherence in a degenerate two-level atomic ensemble:\\
for a transition $F_e=0\leftrightarrow F_g=1$}
\author{Ying Gu$^*$, Qingqing Sun, and Qihuang Gong}
\address{State Key Laboratory for Mesoscopic Physics, Department of Physics,%
\\
Peking University, Beijing 100871, China\\
$^*$e-mail: ygu@pku.edu.cn}
\maketitle

\begin{abstract}
For a transition $F_e=0\leftrightarrow F_g=1$ driven by a linearly polarized
light and probed by a circularly light, quantum coherence effects are
investigated. Due to the coherence between the drive Rabi frequency and
Zeeman splitting, electromagnetically induced transparency,
electromagnetically induced absorption, and the transition from positive to
negative dispersion are obtained, as well as the populations coherently
oscillating in a wide spectral region. At the zero pump-probe detuning, the
subluminal and superluminal light propagation is predicted. Finally,
coherent population trapping states are not highly sensitive to the
refraction and absorption in such ensemble.
\end{abstract}


\vskip 5mm

\section{Introduction}

Quantum coherence or interference effectively influences the property of
refraction and absorption in the phase-coherent atomic ensemble\cite{1}.
When laser resonantly interacts with atoms, the coherence can lead to
electromagnetically induced transparency (EIT)\cite{2}, electromagnetically
induced absorption (EIA) \cite{3,4}, coherent population trapping (CPT)\cite%
{5} and the enhanced refractive index effect\cite{6}. These phenomena have
been widely studied in the typical $\Lambda $- and $V$-type three-level
atomic systems. Recently, the coherently prepared degenerate atomic ensemble
attracts great interest. The CPT in multilevel dark states was
experimentally explored in the ground hyperfine states of $^{87}$Rb atoms%
\cite{7}. EIT and EIA have been observed in the degenerate systems due to
the atomic coherence among Zeeman sublevels belonging to the same hyperfine
splitting\cite{3,8}. In this paper, we use the optical Bloch equations to
study quantum coherence effects induced by the competition between the
coupling field and Zeeman splitting in a degenerate two-level atomic
ensemble.

Since quantum coherence greatly enhances the refractive index and induces
the transparency, the study on the light propagation in a superluminal and
subluminal group velocity is ignited. Wang et al. realized the superluminal
and negative group velocity propagation of light pulses by using the
lossless linear anomalous dispersion between two closely spaced gain lines%
\cite{9}. The experiments have shown that light pulses were greatly slowed
down, halted, and restarted in a coherently prepared atomic media\cite%
{10,11,12,13,14}. Theoretically, by changing the intensity of the lower
level coupling field in a $\Lambda $-type ensemble\cite{15} and adjusting
the phases of two weak optical fields in a $V$-type ensemble\cite{16}, the
group velocity of a light pulse is controlled, so the transition from
subluminal to superluminal light propagation is predicted. In a coherently
prepared degenerate two-level atomic system, steep anomalous and normal
dispersion were observed\cite{8}. The degenerate systems should show more
ability to adjust the sign of dispersion due to the coherence between the
coupling field and Zeeman splitting. Motivated by that, in this work, we
also investigate the positive and negative dispersion in a degenerate
two-level ensemble.

In the following, the simplest degenerate two-level atomic system is
considered. For a transition $F_{e}=0\leftrightarrow F_{g}=1$ driven by a
circularly polarized light with frequency $\omega _{c}$, positive and
negative dispersions of a circular probe light with $\omega _{p}$ are
investigated\cite{17}. At this case, it is still a $\Lambda $-type system
but with a population decay to the state $M_{F_{g}}=0$. Instead of adding a
circularly polarized pumping light, we drive the atomic system by a linear
polarized $\pi $ light with frequency $\omega _{c}$ and probe it by a left
and right circularly polarized $\sigma _{\pm }$ light with $\omega _{p}$.
Now the system becomes a double $\Lambda $-type one. In the presence of the
magnetic field, EIT, EIA, and the transition from positive to negative
dispersion are obtained due to the coherence between the drive Rabi
frequency and Zeeman splitting. Moreover, when the drive Rabi frequency $%
|V_{c}|$ equals to $\omega _{1-1}$ (where $\omega _{1-1}=\omega _{1}-\omega
_{-1}$ and the frequencies $\omega _{1}$ and $\omega _{-1}$ correspond to
the Zeeman sublevels $M_{F_{g}}=1$ and $M_{F_{g}}=-1$), we observe that
populations $\rho _{11}$ and $\rho _{-1-1}$ oscillate at about $0.5$ with
two-cycle. At the zero detuning, $\delta =\omega _{p}-\omega _{c}=0$, the
subluminal and superluminal light propagation is predicted due to the steep
normal and anomalous dispersion. In contrast, when $|V_{c}|<<\omega _{1-1}$
or $|V_{c}|>>\omega _{1-1}$, the coherence between the coupling field and
Zeeman splitting becomes weak, the results of degenerate two-level systems
appear, and only EIT and positive dispersion are observed. Finally, we find
that CPT is not highly sensitive to the refraction and absorption in such
two-level systems.

The paper is organized as follows. In the next section, we set up the model
and present its optical Bloch equations, which can be expressed as a set of
linear equations of Fourier amplitudes. In Section III, we solve these
linear equations numerically and discuss the quantum coherence effects due
to the competition between the coupling field and Zeeman splitting. Finally,
we summary the main results in Section IV.

\section{Optical Bloch equations}

Consider the simplest degenerate two-level system. As shown in Fig. 1, a $%
F_{e}=0\leftrightarrow F_{g}=1$ transition is set up. This transition can be
experimentally realized by using the $4f^{6}6s^{2}$ $^{7}F_{1}%
\leftrightarrow $ $4f^{6}6s6p$ $^{5}D_{0}$ transition of Sm\cite{18,19} and
the $2p^{5}3s$ $^{3}P_{1}\leftrightarrow 2p^{5}3p$ $^{3}P_{0}$ transition of
Ne$^{\ast }$\cite{20}. The atoms are driven by a linear polarized $\pi $
light with frequency $\omega _{c}$ and probed by a left and right circularly
polarized light $\sigma _{\pm }$ with frequency $\omega _{p}$, where $\pi $
light interacts with the transition $M_{F_{e}}=0\leftrightarrow M_{F_{g}}=0$
and $\sigma \pm $ component of circular light interacts with the transition $%
M_{F_{g}}=\mp 1\leftrightarrow M_{F_{e}}=0$. In the presence of magnetic
field, the system is a closed four-level one, but can be looked as a double $%
\Lambda $-type ensemble. All transitions have the same magnitude of dipole
moments, but with the different sign, i.e. $\mu _{e1}=\mu _{e-1}=-\mu
_{e0}=\mu $. Considering the decay of the atomic levels due to the
spontaneous emission and the collisions that result in dephasing of the
coherences and exchange of population between the ground state levels, in
the rotating-wave approximations, we write the optical Bloch equations as:

\begin{equation}
\dot{\rho}_{ee}=\frac{i}{\hbar }(V_{e-1}\rho _{-1e}-\rho _{e-1}V_{-1e})+%
\frac{i}{\hbar }(V_{e1}\rho _{1e}-\rho _{e1}V_{1e})+\frac{i}{\hbar }%
(V_{e0}\rho _{0e}-\rho _{e0}V_{0e})-(\Gamma _{e-1}+\Gamma _{e1}+\Gamma
_{e0})\rho _{ee}
\end{equation}

\begin{equation}
\dot{\rho}_{-1-1}=-\frac{i}{\hbar }(V_{e-1}\rho _{-1e}-\rho
_{e-1}V_{-1e})-(\Gamma _{-10}+\Gamma _{-11})\rho _{-1-1}+\Gamma _{e-1}\rho
_{ee}+\Gamma _{0-1}\rho _{00}+\Gamma _{1-1}\rho _{11}
\end{equation}

\begin{equation}
\dot{\rho}_{11}=-\frac{i}{\hbar }(V_{e1}\rho _{1e}-\rho _{e1}V_{1e})-(\Gamma
_{10}+\Gamma _{1-1})\rho _{11}+\Gamma _{e1}\rho _{ee}+\Gamma _{01}\rho
_{00}+\Gamma _{-11}\rho _{-1-1}
\end{equation}

\begin{equation}
\dot{\rho}_{00}=-\frac{i}{\hbar }(V_{e0}\rho _{0e}-\rho _{e0}V_{0e})-(\Gamma
_{0-1}+\Gamma _{01})\rho _{00}+\Gamma _{e0}\rho _{ee}+\Gamma _{-10}\rho
_{-1-1}+\Gamma _{10}\rho _{11}
\end{equation}

\begin{equation}
\dot{\rho}_{e-1}=-\frac{i}{\hbar }[V_{e-1}(\rho _{ee}-\rho
_{-1-1})-V_{e1}\rho _{1-1}-V_{e0}\rho _{0-1}]-(i\omega _{e-1}+\gamma
_{e-1})\rho _{e-1}
\end{equation}

\begin{equation}
\dot{\rho}_{e1}=-\frac{i}{\hbar }[V_{e1}(\rho _{ee}-\rho _{11})-V_{e-1}\rho
_{-11}-V_{e0}\rho _{01}]-(i\omega _{e1}+\gamma _{e1})\rho _{e1}
\end{equation}

\begin{equation}
\dot{\rho}_{e0}=-\frac{i}{\hbar }[V_{e0}(\rho _{ee}-\rho _{00})-V_{e-1}\rho
_{-10}-V_{e1}\rho _{10}]-(i\omega _{e0}+\gamma _{e0})\rho _{e0}
\end{equation}

\begin{equation}
\dot{\rho}_{1-1}=-\frac{i}{\hbar }(V_{e-1}\rho _{1e}-\rho
_{e-1}V_{1e})-(i\omega _{1-1}+\gamma _{1-1})\rho _{1-1}
\end{equation}

\begin{equation}
\dot{\rho}_{10}=-\frac{i}{\hbar }(V_{e0}\rho _{1e}-V_{1e}\rho
_{e0})-(i\omega _{10}+\gamma _{10})\rho _{10}
\end{equation}

\begin{equation}
\dot{\rho}_{-10}=-\frac{i}{\hbar }(V_{e0}\rho _{-1e}-V_{-1e}\rho
_{e0})-(i\omega _{-10}+\gamma _{-10})\rho _{-10}
\end{equation}

\begin{equation}
\dot{\rho}_{ij}=\dot{\rho}_{ji}^{\ast }.
\end{equation}%
Ignoring the effect of spatial amplitude, the interacting energy $V_{ei}$
for the $\left| e\right\rangle \rightarrow \left| i\right\rangle $
transition can be expressed as $V_{e1}=V_{e-1}=\hbar V_{p}(\omega
_{p})e^{-iw_{p}t}$ and $V_{e0}=\hbar V_{c}(\omega _{c})e^{-iw_{c}t}$. Here
the magnitude values of $2V_{p}(\omega _{p})=\mu E_{p}/2^{1/2}\hbar $ and $%
2V_{c}(\omega _{c})=-\mu E_{c}/2^{1/2}\hbar $ are defined as the probe Rabi
frequency and drive Rabi frequency respectively. Let $\omega _{ij}=$ $\omega
_{i}-$ $\omega _{j}$ and $\omega _{1-1}=2g\mu _{b}B/\hbar $ denotes the
Raman detuning induced by the static magnetic field of strength $B$ applied
in the atoms, where $g$ is the Lande-factor and $\mu _{b}$ the Bohr
magneton. $\Gamma _{ij\text{ \ }}$is the decay rate from state $\left|
i\right\rangle $ to $\left| j\right\rangle $. Here we let $\Gamma
_{e1}=\Gamma _{e-1}=\Gamma _{e0}=\Gamma .$ The decay rate among three lower
levels is small and same, i.e., $\Gamma _{ij}=\Gamma _{0}$ for $%
i,j=1,0,-1(i\neq j)$. Ignoring the rate of dephasing collisions, $\gamma
_{e-1}=\gamma _{e0}=\gamma _{e1}=\gamma =(3\Gamma +2\Gamma _{0})/2$ and $%
\gamma _{1-1}=\gamma _{10}=\gamma _{-10}=\gamma _{0}=2\Gamma _{0}$.

We treat the drive field to all orders in $E_{c}$ and the probe field to the
first order in $E_{p}$, then $\rho _{ei}$ with $i=1,0,-1$ is dominated by
three main frequencies: $\omega _{c},\omega _{p},2\omega _{c}-\omega _{p}$.
In terms of the Fourier amplitudes $\rho _{ei}(\omega _{i})$, $\rho _{ei}$
can be expressed as $\rho _{ei}=\rho _{ei}(\omega _{c})e^{-i\omega
_{c}t}+\rho _{ei}(\omega _{p})e^{-i\omega _{p}t}+\rho _{ei}(2\omega
_{c}-\omega _{p})e^{-i(2\omega _{c}-\omega _{p})t}$. The system is assumed
to satisfy the relation: $\rho _{ee}+\rho _{11}+\rho _{00}+\rho _{-1-1}=1$.
In the steady state, we obtain a set up closed linear equations for the
Fourier amplitudes\cite{21,22}:

\begin{equation}
3i\Gamma \rho _{ee}^{dc}=[V_{p}^{\ast }\rho _{e-1}(\omega _{p})-V_{p}\rho
_{-1e}(-\omega _{p})]+[V_{p}^{\ast }\rho _{e1}(\omega _{p})-V_{p}\rho
_{1e}(-\omega _{p})]+[V_{c}^{\ast }\rho _{e0}(\omega _{c})-V_{c}\rho
_{0e}(-\omega _{c})]
\end{equation}%
\begin{equation}
3i\Gamma _{0}\rho _{-1-1}^{dc}=-[V_{p}^{\ast }\rho _{e-1}(\omega
_{p})-V_{p}\rho _{-1e}(-\omega _{p})]+i(\Gamma -\Gamma _{0})\rho
_{ee}^{dc}+i\Gamma _{0}
\end{equation}%
\begin{equation}
3i\Gamma _{0}\rho _{11}^{dc}=-[V_{p}^{\ast }\rho _{e1}(\omega
_{p})-V_{p}\rho _{1e}(-\omega _{p})]+i(\Gamma -\Gamma _{0})\rho
_{ee}^{dc}+i\Gamma _{0}
\end{equation}%
\begin{equation}
(\omega _{-11}-i\gamma _{0})\rho _{-11}^{dc}=[V_{p}^{\ast }\rho _{e1}(\omega
_{p})-V_{p}\rho _{-1e}(-\omega _{p})]
\end{equation}%
\begin{equation}
(\omega _{c}-\omega _{e0}+i\gamma )\rho _{e0}(\omega _{c})=V_{c}(2\rho
_{ee}^{dc}+\rho _{-1-1}^{dc}+\rho _{11}^{dc}-1)-V_{p}[\rho _{-10}(\omega
_{c}-\omega _{p})+\rho _{10}(\omega _{c}-\omega _{p})]
\end{equation}%
\begin{equation}
(\omega _{p}-\omega _{e-1}+i\gamma )\rho _{e-1}(\omega _{p})=-V_{p}(\rho
_{-1-1}^{dc}-\rho _{ee}^{dc}+\rho _{1-1}^{dc})-V_{c}\rho _{0-1}(\omega
_{p}-\omega _{c})
\end{equation}%
\begin{equation}
(\omega _{p}-\omega _{e1}+i\gamma )\rho _{e1}(\omega _{p})=-V_{p}(\rho
_{11}^{dc}-\rho _{ee}^{dc}+\rho _{-11}^{dc})-V_{c}\rho _{01}(\omega
_{p}-\omega _{c})
\end{equation}%
\begin{equation}
(\omega _{c}-\omega _{p}-\omega _{10}+i\gamma _{0})\rho _{10}(\omega
_{c}-\omega _{p})=-[V_{p}^{\ast }\rho _{e0}(\omega _{c})-V_{c}\rho
_{1e}(-\omega _{p})]
\end{equation}%
\begin{equation}
(\omega _{c}-\omega _{p}-\omega _{-10}+i\gamma _{0})\rho _{-10}(\omega
_{c}-\omega _{p})=-[V_{p}^{\ast }\rho _{e0}(\omega _{c})-V_{c}\rho
_{-1e}(-\omega _{p})]
\end{equation}%
\begin{equation}
\rho _{ij}(\omega _{k})=\rho _{ji}^{\ast }(-\omega _{k})
\end{equation}%
Taking account of Eq. (21), we are readily to solve these linear equations.

Since both transitions $M_{F_{g}}=1\leftrightarrow M_{F_{e}}=0$ and $%
M_{F_{g}}=-1\leftrightarrow M_{J_{e}}=0$ have the same dipole moment and
transverse and logitudinal decay rates, the probe refractive index and
absorption are proportional to the real and imaginary part of the
susceptibility, i.e., $\chi (\omega _{p})\varpropto \lbrack \rho
_{e1}(\omega _{p})+\rho _{e-1}(\omega _{p})]/(V_{2}/\gamma ).$ Because the
dispersion $D$ is proportional to $d[%
\mathop{\rm Re}%
(\chi (\omega _{p}))]/d(\omega _{p})$, at the zero detuning, $\delta =0$,
the dispersion can be written as $D=d\{%
\mathop{\rm Re}%
[\rho _{e1}(\omega _{p})+\rho _{e-1}(\omega _{p})]/(V_{2}/\gamma
)\}/d(\delta /\gamma )$. In the steep dispersion, ignoring the dispersion of
group velocity, we have\cite{23}

\begin{equation}
\frac{c}{V_{g}}=1+2\pi 
\mathop{\rm Re}%
\chi (\omega _{p})+2\pi \omega _{p}%
\mathop{\rm Re}%
(\frac{\partial \chi }{\partial \omega _{p}}).
\end{equation}%
In the real atomic system, $c/V_{g}=1+\Omega D(\delta =0),$where $\Omega
=\pi \omega _{p}N\mu _{e1}^{2}/\gamma ^{2}\hbar $\cite{16}. For a positive
dispersion, i.e. $D>0$, $V_{g}<c$, which is a subluminal light propagation.
While, for a negative dispersion, i.e. $D<0$, $V_{g}>c$, a superluminal
light propagation.

\section{Numerical results and discussions}

To model the system shown in Fig. 1, we first define the parameters in the
Bloch equations, $\gamma =1.0$, $\Gamma _{0}=0.001\gamma $, $\Gamma
=2(\gamma -\Gamma _{0})/3$ and $\gamma _{0}=2\Gamma _{0}$, which are
referred to Ref. [17,21]. We let $\omega _{c}\simeq \omega _{0}$. Here the
Zeeman splitting $\omega _{1-1\text{ }}$ between $M_{F_{g}}=1$ and $%
M_{F_{g}}=-1$ is set up $5.0$, so that $\omega _{10\text{ }}=$ $\omega _{0-1%
\text{ }}=2.5$. To satisfy that the probe Rabi frequency $2|V_{p}|$ is great
smaller than the drive Rabi frequency $2|V_{c}|$, we always let $%
|V_{p}|=0.01|V_{c}|$. For simplicity, we set $V_{c}$ and $V_{p}$ real. We
plot the real and imaginary parts of $[\rho _{e1}(\omega _{p})+\rho
_{e-1}(\omega _{p})]/(V_{2}/\gamma )$, which can be used to represent the
refraction and absorption.

In order to study the refraction and absorption in a transition $%
F_{g}=1\leftrightarrow F_{e}=0$, we first consider the quantum coherence
effects of a degenerate two-level atomic ensemble. It is a superposition of
two $\Lambda $-type systems: $M_{F_{g}}=-1\leftrightarrow
F_{e}=0\leftrightarrow M_{F_{g}}=0$ and $M_{F_{g}}=1\leftrightarrow
F_{e}=0\leftrightarrow M_{F_{g}}=0$, where the state $M_{F_{g}}=0$ shares
the common transition: $F_{e}=0\leftrightarrow M_{F_{g}}=0$. The same
results as what found in a typical $\Lambda $-type system, EIT and positive
dispersion are shown as in Fig. 2. When the pump-probe detuning $\delta $ is
changed at a large spectral region, the populations are almost trapped in
the two ground states $M_{F_{g}}=\pm 1$, namely, $\rho _{11}=\rho
_{-1-1}=0.5 $. Note here EIT, positive dispersion and CPT effects are purely
induced by the linearly polarized coupling $\pi $ light. In the presence of
the magnetic field, the degenerate state $F_{g}=1$ with the energy level $%
\hbar \omega _{0}$ splits into three ones: $M_{F_{g}}=\pm 1$ with $\hbar
\omega _{0}\pm g\mu _{b}B$ and $M_{F_{g}}=0$ with $\hbar \omega _{0}$. When
the intensity of the drive field is varied, due to the existence of Zeeman
splitting, it will experience a great changes in the refraction, absorption
and CPT. The subsequent numerical calculations are performed to discuss the
quantum coherence effects, which are induced by the competition between the
drive Rabi frequency and Zeeman splitting.

It is known that in many ensembles the intensity of coupling field can be
used to control the quantum coherence effects\cite{8,15,17,21}. In the
subsequent figures, by means of changing the intensity of the coupling
field, the quantum coherence effects between the drive Rabi frequency and
the Zeeman splitting are shown. When $|V_{c}|$ is very small compared with
the Zeeman splitting $\omega _{10\text{ }}=2.5$, i.e., $|V_{c}|=1.0<<\omega
_{10\text{ }},$the existence of the coupling field will not be enough to
influence the properties of the ensemble. As shown in Fig. 3, at zero
detuning, EIT and positive dispersion are found. The populations $\rho _{11}$
and $\rho _{-1-1}$ are little deviated from $0.5$ at the small spectral
region $\delta \in \lbrack -5,5]$ due to the existence of the Zeeman
splitting. When $|V_{c}|$ equals to Zeeman splitting $\omega _{10\text{ }}$,
due to the coherence between $|V_{c}|$ and Zeeman splitting, as shown in
Fig. 4, not only EIA occurs, but also the dispersion is exhibited to be
negative. Simultaneously, there are larger changes in CPT states comparing
with the previous case. When we continue to increase $|V_{c}|$ , let it be
two times of Zeeman splitting, i.e. $V_{c}=5.0$, we find that EIT and
positive dispersion occurs at three frequencies: $\omega _{p}=\omega _{c},$
and $\omega _{p}=\omega _{c}\pm 5.0$. These results are shown in Fig. 5. The
corresponding populations $\rho _{11}$ and $\rho _{-1-1}$ oscillate at $0.5$
in a wide spectral region of the detuning $\delta $. The cycle of
oscillation is $2$. This phenomenon results from the coherence effect
between the drive filed and the Zeeman splitting\cite{24}. This kind of
coherence has been discussed in the Ref. [24], while the oscillation of
populations was not mentioned. At last, when the drive Rabi frequency $%
|V_{c}|$ is great larger than the Zeeman splitting, i.e., $%
|V_{c}|=10.0>>\omega _{10\text{ }},$the coupling field dominates the
ensemble, so the results of a degenerate system appear. As shown in Fig. 6,
EIT and positive dispersion are exhibited in a wide spectral region of $%
\delta $ \cite{25}. Therefore, in the presence of the magnetic field, both
EIT and EIA arise due to the coherence between $|V_{c}|$ and Zeeman
splitting.

Here positive and negative dispersions at the zero detuning, $\delta =0,$
are emphasized. We observe the transition from positive, via negative, to
positive dispersion from the features of refraction in Figures 3, 4 and 5.
Since the group velocity $V_{g}$ can be expressed as $c/V_{g}=1+\Omega
D(\delta =0)$\cite{16}, the transition from subluminal to superluminal light
propagation is predicted. We have noted this transition in a $\Lambda $-type
ensemble by changing the intensity of the lower level coupling field\cite{15}
and in a $V$-type ensemble by adjusting the phases of two weak optical fields%
\cite{16}. In the presence of magnetic field, the ensemble becomes a real
four-level system, or a superposition of two $\Lambda $-type ensembles. Due
to the existence of Zeeman sublevel $M_{F_{g}}=0$, the system becomes more
adjustable, so that the changing of $V_{c}$ can lead to the positive and
negative dispersions. Therefore this kind of system is suitable to unify the
subluminal and superluminal light propagation.

CPT states in the excited level and Zeeman sublevels are shown in these
figures. In all figures, at $\delta =0$, the populations $\rho _{11}$ and $%
\rho _{-1-1}$ approximately equal to $0.5$ while $\rho _{ee}$ and $\rho
_{00} $ are approximately zero. The pump effect of a linearly polarized
light results in populations trapping in the Zeeman sublevels $M_{F_{g}}=1$
and $M_{F_{g}}=-1$. In the ensemble driven and probed by a circular
polarized light\cite{17}, the quantum coherent result is different, in that,
at the detuning center the ratio $\frac{\rho _{ee}}{\rho _{11}+\rho _{-1-1}}$
is large and can reach the maximum $\frac{1}{2}$. Finally, we find that CPT
states are not sensitively related to the refraction and absorption in such
ensemble.

The opposite case is complemented, where the atomic system is driven by a
left and right circularly polarized $\sigma _{\pm }$ light with $\omega
_{c\pm }$ and probed by a linear polarized $\pi $ light with $\omega _{p}$.
Analogous to the system driven by a $\pi $ light and probed by a $\sigma
_{\pm }$ light, it is also a double $\Lambda $-type ensemble. At the
degenerate case, it should exhibit the same property as what found in
previous, i.e., EIT and positive dispersion, which have been verified by
further numerical calculations. However, when the magnetic field is added to
the ensemble, there exist more than three dominant frequencies: $\omega
_{c+},$ $\omega _{c-},$ $\omega _{p},2\omega _{c+}-\omega _{p},2\omega
_{c-}-\omega _{p}$ and so on, so that it is very difficult to obtain the
Fourier amplitudes. For this system, it should exhibit the same quantum
coherence effects as what discussed in the previous part of the paper. In
ref. [17], a transition $F_{g}=1\leftrightarrow F_{e}=0$ driven and probed
by a circularly polarized light is investigated. At that case, though the
existence of the state $M_{F_{g}}=0$ the system is still a $\Lambda $-type
one, so only positive or negative dispersion\ and EIT are found. Finally,
for a transition $F_{e}=0\leftrightarrow F_{g}=1$ driven and probed by a
linearly polarized light, only transition $M_{F_{e}}=0$ $\longleftrightarrow
M_{F_{g}}=0$ exists, it corresponds to a two-level atomic ensemble.

\section{Conclusion}

In summary, we investigate the quantum coherence effects in a degenerate
two-level atomic system. For a transition $F_{e}=0\leftrightarrow F_{g}=1$
driven by a linearly polarized light and probed by a circularly polarized
light, a double $\Lambda $-type ensemble is set up. In the presence of the
magnetic field, due to the coherence between the drive Rabi frequency and
Zeeman splitting, EIT, EIA, and the transition from positive to negative
dispersion are obtained. When the drive Rabi frequency $|V_{c}|$ equals to
Zeeman splitting, we observe that the populations $\rho _{11}$ and $\rho
_{-1-1}$ oscillate at about $0.5$ with two-cycle. At the zero detuning, the
subluminal and superluminal light propagation is predicted due to the steep
normal and anomalous dispersion. While, when drive Rabi frequency is great
larger than or smaller than Zeeman splitting, the action of Zeeman splitting
becomes weak, the results of degenerate two-level systems appear, namely,
only EIT and positive dispersion are observed. Finally, CPT is not highly
sensitive to the refraction and absorption as we expected.

\section{Acknowledgments}

The work was supported by the special fund from Ministry of Science and
Technology, China, the National Key Basic Research Special Foundation
(NKBRSF) under grant no. G1999075207 and National Natural Science Foundation
of China under grant nos. 19884001 and 90101027. Y. G. acknowledges useful
discussions with Professor R. P. Wang.

\begin{figure}[tbp]
\caption{Schematic diagram of energy-level for a transition $%
F_{e}=0\leftrightarrow F_{g}=1$ driven by a linearly polarized $\protect\pi $
light and probed by a circularly polarized $\protect\sigma _{\pm }$ light. }
\label{fig1}
\end{figure}

\begin{figure}[tbp]
\caption{Quantum coherence effects of a degenerate two-level atomic
ensemble. The transition is $F_{e}=0\leftrightarrow F_{g}=1$. Here $%
|V_{c}|=1.0$ and $|V_{p}|=0.01|V_{c}|$. (a) CPT for four populations $%
\protect\rho _{ee},\protect\rho _{11},\protect\rho _{00}$ and $\protect\rho %
_{-1-1}$. The atoms are trapped in the Zeeman sublevels $M_{F_{g}}=1$ and $%
M_{F_{g}}=-1$ with the same probability $\protect\rho _{11}=\protect\rho %
_{-1-1}=0.5$. (b) Real and imaginary parts of $[\protect\rho _{e1}(\protect%
\omega _{p})+\protect\rho _{e-1}(\protect\omega _{p})]/(V_{2}/\protect\gamma %
)$, which are proportional to refraction and absorption. EIT and positive
dispersion are shown at the zero detuning $\protect\delta =0.0$.}
\label{fig2}
\end{figure}

\begin{figure}[tbp]
\caption{ Quantum coherence effects for a transition $F_{e}=0\leftrightarrow
F_{g}=1$ with Zeeman splitting $\protect\omega _{10}=2.5$ and drive Rabi
frequency $|V_{c}|=1.0$. (a) CPT for four populations $\protect\rho _{ee},%
\protect\rho _{11},\protect\rho _{00}$ and $\protect\rho _{-1-1}$. Most
atoms are trapped in the Zeeman sublevels $M_{F_{g}}=1$ and $M_{F_{g}}=-1$.
(b) Real and imaginary parts of $[\protect\rho _{e1}(\protect\omega _{p})+%
\protect\rho _{e-1}(\protect\omega _{p})]/(V_{2}/\protect\gamma )$, which
are proportional to refraction and absorption. EIT and positive dispersion
are found at the zero detuning $\protect\delta =0.0$. }
\label{fig3}
\end{figure}

\begin{figure}[tbp]
\caption{ Quantum coherence effects for a transition $F_{e}=0\leftrightarrow
F_{g}=1$ with Zeeman splitting $\protect\omega _{10}=2.5$ and drive Rabi
frequency $|V_{c}|=2.5$. (a) CPT for four populations $\protect\rho _{ee},%
\protect\rho _{11},\protect\rho _{00}$ and $\protect\rho _{-1-1}$. Most
atoms are trapped in the Zeeman sublevels $M_{F_{g}}=1$ and $M_{F_{g}}=-1$,
but with a little oscillation around $0.5$. (b) Real and imaginary parts of $%
[\protect\rho _{e1}(\protect\omega _{p})+\protect\rho _{e-1}(\protect\omega %
_{p})]/(V_{2}/\protect\gamma )$, which are proportional to refraction and
absorption. Due to the coherence between drive Rabi frequency and Zeeman
splitting, EIA and negative dispersion are exhibited at the zero detuning $%
\protect\delta =0.0$. }
\label{fig4}
\end{figure}

\begin{figure}[tbp]
\caption{ Quantum coherence effects for a transition $F_{e}=0\leftrightarrow
F_{g}=1$ with Zeeman splitting $\protect\omega _{10}=2.5$ and drive Rabi
frequency $|V_{c}|=5.0$. (a) CPT for four populations $\protect\rho _{ee},%
\protect\rho _{11},\protect\rho _{00}$ and $\protect\rho _{-1-1}$. Most
atoms are trapped in the Zeeman sublevels $M_{F_{g}}=1$ and $M_{F_{g}}=-1$,
but with a two-cycle oscillation around $0.5$. (b) Real and imaginary parts
of $[\protect\rho _{e1}(\protect\omega _{p})+\protect\rho _{e-1}(\protect%
\omega _{p})]/(V_{2}/\protect\gamma )$, which are proportional to refraction
and absorption. Due to the coherence between drive Rabi frequency and Zeeman
splitting, EIT and positive dispersion occur at three frequencies: $\protect%
\omega _{p}=\protect\omega _{c},$ and $\protect\omega _{p}=\protect\omega %
_{c}\pm 5.0$. }
\label{fig5}
\end{figure}

\begin{figure}[tbp]
\caption{ Quantum coherence effects for a transition $F_{e}=0\leftrightarrow
F_{g}=1$ with Zeeman splitting $\protect\omega _{10}=2.5$ and drive Rabi
frequency $|V_{c}|=10.0$. (a) CPT for four populations $\protect\rho _{ee},%
\protect\rho _{11},\protect\rho _{00}$ and $\protect\rho _{-1-1}$. Most
atoms are trapped in the Zeeman sublevels $M_{F_{g}}=1$ and $M_{F_{g}}=-1$,
but with a large deviation from $0.5$ far from the pump-probe detuning
center. (b) Real and imaginary parts of $[\protect\rho _{e1}(\protect\omega %
_{p})+\protect\rho _{e-1}(\protect\omega _{p})]/(V_{2}/\protect\gamma )$,
which are proportional to refraction and absorption. The results of a
degenerate ensemble appear, and EIT and positive dispersion are found at a
large spectral region.}
\label{fig.6}
\end{figure}


\begin{references}
\bibitem{1} M. O. Scully and M. S. Zubairy, {\it {Quantum optics}}
(Cambridge Univ. Press) (1997).

\bibitem{2} S. E. Harris, Phys. Today {\bf {50 No. 7}}, 36 (1997).

\bibitem{3} A. Lezama, S. Barreiro, and A. M. Akulshin, Phys. Rev. A {\bf {59%
}}, 4732 (1999).

\bibitem{4} A. M. Akulshin, S. Barreiro, and A. Lezama, Phys. Rev. A {\bf {57%
}}, 2996 (1998).

\bibitem{5} E. Arimondo, in {\it {Progress in Optics}} edited by E.
Wolf(Elsevier Science, Amsterdam) {\bf {35}}, 257 (1996).

\bibitem{6} M. D. Lukin, Scully MO, Welch GR, et al., Laser Physics {\bf 9},
759 (1999).

\bibitem{7} YiFu Zhu, Shijun Wang, and Neil M. Mulchan, Phys. Rev. A {\bf 59}%
, 4005 (1999).

\bibitem{8} A. M. Akulshin, S. Barreiro, and A. Lezama, Phys. Rev. Lett. 
{\bf 83}, 4277 (1999).

\bibitem{9} L. J. Wang, A. Kuzmich, and A. Dogaiu, Nature (London) {\bf 406}%
, 277 (2000).

\bibitem{10} L. V. Hau, S. E. Harris, Z. Dutton, and C. H. Behroozi, Nature
(London) {\bf 397}, 594 (1999).

\bibitem{11} M. D. Lukin and A. Imamoglu, Phys. Rev. Lett. {\bf 84}, 1419
(2000).

\bibitem{12} C. Liu, Z. Dutton, C. H. Behroozi, et al., Nature (London) {\bf %
409}, 490 (2001).

\bibitem{13} D. F. Phillips, A. Fleischhauer, A. Mair, et al., Phys. Rev.
Lett. {\bf 86}, 783 (2001).

\bibitem{14} A. V. Turukhin, V. S. Sudarshanam, M. S. Shahriar, et al.,
Phys. Rev. Lett. {\bf 88}, 023602 (2002).

\bibitem{15} G. S. Agarwal, T. N. Dey, and S. Menon, Phys. Rev. A {\bf 64},
053809 (2001).

\bibitem{16} D. Bortman-Arbiv, A. D. Wilson-Gordon, and H. Friedmann, Phys.
Rev. A {\bf 63}, 043818 (2001).

\bibitem{17} A. D. Wilson-Gordon and H. Friedmann, J. Mod. Opt. {\bf 49},
125 (2002).

\bibitem{18} L. M. Barkov, et al., Opt. Commun. {\bf 70}, 467 (1989).

\bibitem{19} S. A. Diddams, J. C. Diels, and B. Atherton, Phys. Rev. A {\bf %
58}, 2252 (1998).

\bibitem{20} K. Bergmann, H. Theuer, and B. W. Shore, Rev. Mod. Phys. {\bf 70%
}, 1003 (1998).

\bibitem{21} A. D. Wilson-Gordon, Phys. Rev. A {\bf 48}, 4639 (1993).

\bibitem{22} R. W. Boyd, et al., Phys. Rev. A {\bf 24}, 411 (1981).

\bibitem{23} S. E. Harris, et al., Phys. Rev. A {\bf 46}, R29 (1992).

\bibitem{24} S. Menon and G. S. Agarwal, Phys. Rev. A {\bf 59}, 740 (1999).

\bibitem{25} A. Lezama, S. Barreiro, A. Lipsich, and A. M. Akulshin, Phys.
Rev. A {\bf 61}, 013801 (1999).
\end{references}
\end{document}